\documentclass[a4paper,twocolumn,english,pre]{revtex4}
\usepackage[T1]{fontenc}
\usepackage[latin1]{inputenc}
\usepackage{amsmath}
\usepackage{graphicx}
\usepackage{amssymb}

\makeatletter
\usepackage{babel}
\makeatother
\begin{document}

\title{Long-Range Order and Interactions of Macroscopic Objects in Polar
Liquids. }

\author{P.O. Fedichev$^{1,2}$, L.I. Men'shikov$^{2}$}

\address{$^{1)}$Quantum Pharmaceuticals Ltd, Usievicha str 8-131, 125319,
Moscow, Russian Federation, e-mail: peter.fedichev@q-pharm.com and }

\address{$^{2)}$RRC Kurchatov Institute, Kurchatov Square 1, 123182, Moscow,
Russian Federation }

\begin{abstract}
We develop a phenomenological vector model of polar liquids capable
to describe aqueous interactions of macroscopic bodies. It is shown
that a strong, long-range and orientationally dependent interaction
between macroscopic objects appears as a result of competition between
short-range (hydrogen bonding) and the long-range dipole-dipole interactions
of the solvent molecules. Spontaneous polarization of molecular dipoles
next to a hydrophobic boundaries leads to formation of globally ordered
network of hydrogen-bonded molecules with ferroelectric properties.
The proposed vector model naturally describes topological excitations
on the solute boundaries and can be used to explain the hydrogen bonds
networks and order-disorder phase transitions in the hydration water
layer. 
\end{abstract}
\maketitle

\section{Introduction.}

Interactions of macroscopic bodies in aqueous environments is a fundamentally
important problem in physics, chemistry, structural biology and in
silico drug design. The main theoretical difficulty arises both from
strong long-range dipole-dipole interactions of molecular dipole moments
and complicated nature of short-range interaction between water molecules.
For practical purposes, e.g. calculation of binding free energy (inhibition
constant) of a protein-ligand complex, protonation states predictions
for biomacromolecules, physics of membranes etc, the ultimate way
to quantitatively account for solvation effects is the Free Energy
Perturbation method (FEP) based on a direct modelling of both the
solute and the solvent molecules, molecular dynamics (MD). One of
the main advantages of MD is a possibility to include the surrounding
water molecules into the calculation directly (simulations with explicit
water). In fact this is so far one of the most accurate ways to compute
the solvation and the binding energies \cite{ModernMD}. The limitation
of the method comes together with its strength: to do it one should
compute coordinates and velocities of very large number of atoms with
time step $\tau_{S}\sim10^{-15}s.$ To allow the aqueous environment
to relax fully, the averaging needs to be performed on a sufficiently
long time span to include at least the life-time of hydrogen bonds
($\sim10^{-9}s$), and even the much larger relaxation and rearranging
time of macroscopic water clusters (up to $\sim10^{-6}s$, see discussion
below). For modern computers it is possible to follow the evolution
of macromolecules in water environment up to times $\sim10^{-7}s$,
which may not be sufficient to relax the surrounding water molecules
in some situations. Although such calculations can be very accurate,
realistic applications require enormous computational resources. 

Polar liquids such as water are characterized by large values of static
dielectric constants $\epsilon\gg1$. That is why to a large extent
solvation theory can be reduced to a macroscopic electrostatics with
a solvent being modelled as a high-$\epsilon$ media and the solute
treated as being emerged in a low-$\epsilon$ cavity. This approach
dates back to early Born's papers and can be very successful for quantitative
predictions of solvation energies of small molecules or biomolecules
interactions \cite{Karplus,Onufriev}. On other hand the interactions
of sufficiently small neutral objects is dominated by hydrophobic
effect, which according to \cite{Chandler,Lum,Wolde} originates from
short-range interactions of the solvent molecules. Density functional
models \cite{Chandler,Lum,Wolde} successfully explain the hydrophobic
interactions on molecular scales, though inclusion of electrostatic
interactions appears to be a difficult task. Moreover, understanding
of a number of important observations revealed from MD simulations
must require complete inclusion of long-range interactions and thus
a more sophisticated approach. Among the examples are the arrangement
of the water molecules dipole moments parallel to the hydrophobic
surface \cite{MD1,MD2,MD3}, the vortex-like structures of molecular
dipole moments networks and dipole-bridges near and between the solvated
objects \cite{Higo}. 

In this Letter we introduce a continuous vector model of a polar liquid
capable both the short- and the long-range features of a polar liquid
in a single theoretical framework. The model can be applied to aqueous
interactions of macroscopic bodies of various shapes and charges.
It is shown that the competition between the short range (hydrogen
bonding) and the long-range dipole-dipole interactions of the solvent
molecules leads to appearance of strong, long range and orientationally
dependent interactions between macroscopic objects. A polar liquid
itself is characterized by a complicated fluctuating thermal state,
ordered at sufficiently short scales within a single domain, and completely
disordered at larger distances. This physical picture has far reaching
consequences, especially at solvent-solute surfaces, where the ferroelectric
film of solvent molecules is formed. With temperature increasing fluctuations
of molecular dipole moments can lead to topological phase transitions
depending on hydrophobic properties of the interface. We argue that
the dynamics of macroscopic topological excitations on a surface of
the liquid may be a physical picture behind the percolation transition
of the hydrogen bonding network observed in MD simulations \cite{Oleinikova1,Oleinikova2,Oleinikova3}.

The manuscript is organized as follows. After the introduction in
Section 2 we discuss the basic assumptions and formulate the vector
model of polar liquids mathematically. In following Section we discuss
the bulk properties and the correlation functions of the liquid within
the model. The results of the discussion can be used to calculate
various asymptotic forms of interactions between point-like and macroscopic
bodies of different shapes and charged states in a solute. At last,
in Section 4 we address the properties of hydration layers of molecular
dipoles on macroscopic surfaces with the emphasis on topological properties
of hydrogen bonds networks and possible phase transitions.

\section{The model.}

Polar liquids are similar to ferroelectrics. Having this analogy in
mind it is possible to develop a vector field theory in which the
liquid is described by a local mean value of the molecular polarization
vector \begin{equation}
\mathbf{s}(\mathbf{r})=\langle\mathbf{d}\rangle/d_{0}\label{MeanPolarizationVector}\end{equation}
Here $d_{0}$ is the static dipole momentum of a single molecule,
and the averaging of molecular dipole momentums $\mathbf{d}$ is performed
over a small but macroscopic volume of the liquid containing macroscopic
number of molecules. Mathematical structure of theory of ferroelectrics
\cite{Ginzburg} is analogous to the phenomenological Landau-Ginzburg
theory for ferromagnetics, which deals with the average value of atomic
magnetic moment. All essential properties of a polar liquid are the
result of competition of two opposite effects, two types of interaction:
the long-range electrostatics (dipole-dipole interaction) $\Omega_{dd}$
and the short-range intermolecular potential, $\Omega_{H}$, responsible
for hydrogen bonds (H-bonds) formation. At room temperatures quantum-mechanical
effects dominate in $\Omega_{H}$ . As for $\Omega_{dd}$ , it has
pure classical nature. As the Earnshaw's theorem says, the equilibrium
state is impossible for the system of classical charges \cite{Sivukhin,Stratton}.
It means that $\Omega_{dd}$ leads to chaotization of molecular dipole
momenta orientations, and hence the ground state of a polar liquid
is a disordered state at least on a large scale. At smaller distances
however, the molecules tend to form maximal possible number of H-bonds,
and thus the H-bond interactions order the molecular dipole moments. 

Below we employ the following expression for the free energy of the
liquid:\begin{eqnarray}
\Omega(\mathbf{s(}r\mathbf{)}) & = & \Omega_{H}+\Omega_{dd}-\int dVP_{0}\mathbf{sE_{\mathbf{e}}},\label{TotalFreeEnergy}\end{eqnarray}
where $\mathbf{E}_{e}$ is the external electric field, $P_{0}=nd_{0}$,
and $n$ is the density of molecules in a fluid. The general form
of two first terms in (\ref{TotalFreeEnergy}) is:\begin{equation}
\Omega_{H}+\Omega_{dd}=\frac{1}{2}\sum_{\alpha,\beta}\int dVdV's_{\alpha}(\mathbf{r})s_{\beta}(\mathbf{r'})M_{\alpha\beta}(\mathbf{r}-\mathbf{r'}).\label{GeneralFormOfIntermolecularInteraction}\end{equation}
Here $\alpha,\beta=x,y,z$ enumerate the Cartesian components of the
vector $\mathbf{s}$. In a free liquid, $\mathbf{E}_{e}=0$, Eqs.(\ref{TotalFreeEnergy})
and (\ref{GeneralFormOfIntermolecularInteraction}) give the following
expression for the correlation function:\begin{equation}
Q_{\alpha\beta}(\mathbf{r-\mathbf{r'}})=\langle s_{\alpha}(\mathbf{r})s_{\beta}(\mathbf{r'})\rangle=TM_{\alpha\beta}^{-1}(\mathbf{r}-\mathbf{r'}),\label{CorrelationFunction}\end{equation}
where $T$ is the absolute temperature (in energy units) and the kernel
$M_{\alpha\beta}^{-1}$ is defined as:\begin{equation}
\sum_{\gamma}\int dV_{1}M_{\alpha\gamma}^{-1}(\mathbf{r}-\mathbf{r}_{1})M_{\gamma\beta}(\mathbf{r}_{1}-\mathbf{r'})=\delta_{\alpha\beta}\delta(\mathbf{r}-\mathbf{r'}).\label{DefinitionOfInverseMatrix}\end{equation}
The {}``matrix'' $M_{\alpha\beta}$ can be deduced from the correlation
function taken, e.g. from molecular dynamics calculations \cite{Kolafa}.
In an isotropic liquid the correlation function has the form: \begin{equation}
Q_{\alpha\beta}(\mathbf{r})=a(r)\delta_{\alpha\beta}+b(r)(\hat{r}_{\alpha}\hat{r}_{\beta}-\frac{1}{3}\delta_{\alpha\beta}),\label{GeneralFormOfCorrelationFunction}\end{equation}
where $\hat{\mathbf{r}}=\mathbf{r}/r$. The functions $a(r)$ and
$b(r)$ are related to the following correlation functions: $a(r)=\langle\mathbf{s}(0\mathbf{)s}(\mathbf{r})\rangle/3$,
$b(r)=3(D(r)-a(r))/2$, and $D(r)=\langle(\mathbf{\hat{\mathbf{r}}s}(0))(\hat{\mathbf{r}}\mathbf{s}(\mathbf{r}))\rangle$.
The functions $a(r)$and $D(r)$ can be expressed with the help of
the full pair correlation function $g(r,\theta_{1},\theta_{2},\Phi)$
introduced in \cite{Gray}. Here $\theta_{1},$ $\theta_{2}$, $\phi_{1}$and
$\phi_{2}$ ($\Phi=\phi_{1}-\phi_{2}$) are the polar and azimuthal
angles between vectors of molecular polarities,$\mathbf{s}_{1}\equiv\mathbf{s}(0),$
\textbf{$\mathbf{s}_{2}\equiv\mathbf{s}(\mathbf{r})$,} and the intermolecular
axis \textbf{$\mathbf{r}$}. The correlation function $g$ can be
expanded in spherical harmonics:\begin{equation}
g(r,\theta_{1},\theta_{2},\Phi)=4\pi\sum_{l_{1}l_{2}m}g_{l_{1}l_{2}m}(r)Y_{l_{1}m}(\theta_{1},\phi_{1})Y_{l_{2}-m}(\theta_{2},\phi_{2}).\label{eq:g}\end{equation}
 So that, $g_{000}(r)$ is a usual pair correlation function, which
does not include information about molecular polar vectors. The correlation
function (\ref{eq:g}) gives the probability $dW$ to find two molecules
at a distance $r$ with polarization vectors close to $\mathbf{s}_{1}$,
$\mathbf{s}_{2}$: \[
dW=\frac{g(r,\theta_{1},\theta_{2},\Phi)}{g_{000}(r)}\frac{d\Omega_{1}}{4\pi}\frac{d\Omega_{2}}{4\pi}.\]
Applying all the definitions above we find:\[
a(r)=\frac{1}{9}\left[2g_{111}(r)+g_{110}(r)\right]/g_{000}(r),\]
 \[
b(r)=\frac{1}{3}\left[g_{110}(r)-g_{111}(r)\right]/g_{000}(r).\]
The functions $a$ and $b$are unique characteristics of intermolecular
interactions. 

At sufficiently large scales the free energy functional (\ref{TotalFreeEnergy})
has a few universal features. Since the dipole moments of water molecules
are ordered at small scales by hydrogen bonding, the mean molecular
polarization vector $\mathbf{s}(\mathbf{r})$ is a smooth continuous
function. Therefore, the free energy of the liquid should allow expansion
in powers of the molecular polarization gradients. That is why we
can use the Oseen's like general expression (see e.g. \cite{LandLifshtiz})
for the hydrogen bonding energy:\begin{equation}
\Omega_{H}=P_{0}^{2}\int dV\left(\frac{C}{2}\sum_{\alpha,\beta}\frac{\partial s_{\alpha}}{\partial x_{\beta}}\frac{\partial s_{\beta}}{\partial x_{\alpha}}+\frac{C^{\prime}}{2}(\nabla\mathbf{s})^{2}+V(\mathbf{s}^{2})\right),\label{eq:1}\end{equation}
where $C,C^{\prime}$ are constants, and the scalar function $V(s^{2})$
takes into account short-distance part (other than dipole-dipole interactions)
of the intermolecular interaction potential. The long-range part (dipole-dipole
) of intermolecular interaction energy takes form: \begin{equation}
\Omega_{dd}={\displaystyle \int dV\frac{1}{8\pi}\mathbf{E}_{P}^{2}},\label{eq:Omega_polarization}\end{equation}
so that the full electric field in the liquid, \textbf{$\mathbf{E}$}
is the sum of the polarization and the external fields: \textbf{$\mathbf{E}=\mathbf{E}_{P}+\mathbf{E}_{e}$}.
The polarization electric field $\mathbf{E}_{P}$, satisfies the Poisson
equation, having as a right-hand-side the density of polarization
charges $\rho_{P}=-{\rm div}\mathbf{P}$, where $\mathbf{P}=P_{0}\mathbf{s}$
is the polarization of the liquid. 

In the bulk of a liquid in weak fields ($E\ll P_{0}$) the vector
model defined by Eqs.(\ref{TotalFreeEnergy}),(\ref{eq:1}) and (\ref{eq:Omega_polarization})
can be studied in linearized approximation. First we expand the function
$V(s^{2})$ in powers of $s^{2}$:\begin{equation}
V(s^{2})=\frac{A}{2}s^{2}(\mathbf{r}),\label{VForSmalls}\end{equation}
where $A$ is a dimensionless constant related to long-range correlation
properties of the liquid (see below). Fourier transforming the model
we find that:\[
\Omega=\frac{P_{0}^{2}}{2}\sum_{k,\alpha,\beta}s_{\alpha}s_{\beta}\left((Ck^{2}+A)\delta_{\alpha\beta}+C^{\prime}k_{\alpha}k_{\beta}\right)+\]
\[
+iP_{0}\sum_{k,\alpha}k_{\alpha}s_{\alpha}\phi+\sum_{k}\frac{k^{2}\phi^{2}}{8\pi}+\Omega_{ext},\]
where $k_{\alpha}$ is the wavevector, $\phi$ is the electrostatic
potential. The potential $\Omega_{ext}$ describes the interaction
of the linearized liquid with external objects.

\section{Long range interactions between solutes.}

As seen from Eqs.(\ref{eq:1}),(\ref{eq:Omega_polarization}) and
(\ref{VForSmalls}) polar liquid is naturally characterized by the
two important scales: $L_{T}=\sqrt{C/A}$ and $R_{D}=\sqrt{C/(A+4\pi)}$.
Indeed, on a molecular scale in small regions the dipole moments of
the molecules are correlated and there exist pretty large electric
fields: $\mathbf{E}_{P}\sim P_{0}\sim1\,10^{7}V/cm$. The size of
such polarized region (a domain) is determined by the competition
of dipole-dipole electrostatic and short-range hydrogen bonding forces:
inside the domain $\Omega_{H}\sim\Omega_{dd}$. Since $\Omega_{H}\sim CP_{0}^{2}R_{0}$
and $\Omega_{dd}\sim P_{0}^{2}R_{0}^{3}$, where $R_{0}$ is the size
a polarized region, the scale $R_{D}\sim\sqrt{C}\sim7\textrm{Å}$
(the estimate $C\sim5\,10^{-15}cm^{2}$ can be obtained from the value
of the surface tension) can be considered as a typical size of domain
(the existence of such domains in a bulk of polar liquids was pointed
in \cite{Pauling}). The second scale, $L_{T}$, includes $\sim100$
molecules and describes the correlations between such macroscopic
domains and can be called as superdomain. It is the longest scale
of intermolecular correlations in water (detailed consideration will
be published elsewhere): at larger distances thermal fluctuations
dominate. Superdomains determine a reaction of polar liquid on a weak
uniform static electric field, i.e. its static dielectric constant.
Experimental data for frequency dependence $\epsilon(\omega)$ are
discussed in \cite{WaterDispersion1,WaterDispersion2,Zatsepina}.
At room temperature the transition from the static value $\sim80$
to the universal low value $\sim5$ characterizing the internal rotational
and electronic degrees of freedom of individual water molecules is
spread over a broad frequency range:\begin{equation}
10^{5}\alt\omega\alt10^{7}s^{-1}.\label{FrequencyInterval}\end{equation}
 Such sophisticated behaviour can be described with a help of a simple
Debye model of polarization relaxation (see \cite{ll:volVI,Sivukhin},
for example). In a weak electric field $\mathbf{E}$ the time evolution
of liquid polarization follows the linearized phenomenological kinetic
equation $\mathbf{\dot{P}}=-\left(\mathbf{P}-(\epsilon-1)\mathbf{E}/4\pi\right)/\tau_{0}$,
where $\tau_{0}$ is the Debye relaxation time. The solution gives
the dielectric function $\epsilon(\omega)=\left(\epsilon-i\omega\tau_{0}\right)/\left(1-i\omega\tau_{0}\right)$
characterized indeed by a broad transient region with $\tau_{0}\sim10^{-6}s$.
The latter quantity is much longer than the average life-time of a
single hydrogen bond, $\sim10^{-9}s$, and thus should be related
to water domains rearrangement processes. 

The model (\ref{eq:1}),(\ref{eq:Omega_polarization}) and (\ref{VForSmalls})
is both non-local and non-linear. Full analysis of the solutions requires
numerical simulations and is beyond the scope of this Letter. In what
follows we confine ourselves to the solutions of linearized form of
the equations. In spite of being quite a drastic simplification, this
approach reveals the asymptotic long-range interactions between solute
objects and shows the importance of topological excitations, both
in the bulk and on a surface of the liquid.

\subsection{Interaction between charged objects. Solvation energy of a point
charge.}

Consider first the simplest case of interaction with a number of external
point electric charges $q_{a}$ placed in positions $\mathbf{r}_{a}$:\[
\Omega_{ext}=\int d^{3}r\rho(r)\phi(r),\]
where $\rho(\mathbf{r})=\sum_{a}q_{a}\delta(\mathbf{r}-\mathbf{r}_{a})$.
Then the free energy of the liquid is (setting $C^{\prime}=0$ for
clarity):\begin{equation}
\Omega=\sum_{k}|\rho(\mathbf{k})|^{2}\frac{4\pi}{\epsilon(k)k^{2}},\label{eq:Omega_qq}\end{equation}
where\begin{equation}
\epsilon(k)=\epsilon\frac{R_{D}^{2}k^{2}+1}{L_{T}^{2}k^{2}+1},\label{eq:epsilonK}\end{equation}
is the effective dielectric constant at wavevector $k$, and $\epsilon=L_{T}^{2}/R_{D}^{2}=1+4\pi/A$.
The relation between the material constant $A$ coming from the {}``equation
of state'' (\ref{VForSmalls}) and the macroscopic dielectric susceptibility
$\epsilon$ can be seen from the following observation: For a pair
of point charges separated by a large distance $R$ Eq.(\ref{eq:Omega_qq})
gives the following expression for the interaction energy:\begin{equation}
\Omega(R)=\frac{q_{1}q_{2}}{R}\left[\frac{1}{\epsilon}+(1-\frac{1}{\epsilon})\exp\left(-\frac{R}{R_{D}}\right)\right].\label{InteractionOfPointCharges}\end{equation}
That is why at large distances $R\gg R_{D}$ two charges interact
as in a dielectric media characterized by dielectric constant $\epsilon$.
In a polar liquid $\epsilon\gg1$ and the constant $A\ll1$ (for example
at room temperature in water $\epsilon\approx80$ and $A=4\pi/(\epsilon-1)\approx0.16$). 

The relation between $A$ and $\epsilon$ can also be established
from a well known relation between the asymptotic behavior of linear
response functions of a liquid (correlators) and the dielectric constant
(see e.g. \cite{Ramirez,Kolafa}). Since $A\ll1$, $R_{D}\ll L_{T}\sim15\textrm{Å}$.
Subtracting the vacuum energy of a point charge from Eq.(\ref{eq:Omega_qq})
find the solvation free energy of a point charged object: \begin{equation}
\Omega_{solv}=\sum_{k}|\rho(\mathbf{k})|^{2}\frac{4\pi}{k^{2}}\left(\frac{1}{\epsilon(k)}-1\right)=\left(\frac{1}{\epsilon}-1\right)\frac{q^{2}}{2R_{D}}\label{SolvationEnergyPointCharge}\end{equation}
The result looks pretty much the same as classical Born solvation
theory result. In fact, the solvation energy is localized in the electric
domain of size $k^{-1}\sim R_{D}$ around the charge. In practical
case $R_{D}$ does not exceed much size of an ion (or a charged molecule).
That is why the exact value of ionic Born radius $R_{B}=-q^{2}/2\Omega_{solv}\sim R_{D}$
depends on precise details of microscopic interactions of the ion
and the surrounding water molecules. The results (\ref{InteractionOfPointCharges})
and (\ref{SolvationEnergyPointCharge}) come from the linearized model
and hence are only valid for point-size charges $q\alt q_{C}=CP_{0}$\cite{PieceOfIce}.

\subsection{Interactions of neutral objects. Typical types of liquid polarization.}

The interaction of macroscopic objects is not confined by electrostatic
interaction of the charges. As demonstrated by direct numerical simulations,
macroscopic objects polarize the liquid \cite{MD1,MD2,MD3}. For example,
next to a macroscopic surface the water molecules arrange themselves
such as their dipole moments $\mathbf{d}$ are parallel to the boundary
$(\mathbf{s}\neq0,\,\mathbf{s}\perp\mathbf{n})$ with water molecules
planes being perpendicular to the outer normal $\mathbf{n}$ to the
surface of a body. The explanation is as follows. The bulk water molecule
has $N_{H}\approx4$ hydrogen bonds with others. If the molecular
polarization is orthogonal to the solute body, $\mathbf{d}\parallel\mathbf{n}$,
then $N_{H}\approx2$ , and if $\mathbf{d}\perp\mathbf{n}$, then
$N_{H}\approx3$. Therefore the liquid can lower its free energy by
arranging the molecular dipole moments along the hydrophobic surfaces
(if no energy benefit can be extracted by hydrogen bond formation
with the solute atoms). This simple picture gives the following estimation
for the surface tension coefficient: $\alpha\sim(E_{H}/2)n^{-2/3}\sim P_{0}^{2}R_{D}$,
where $E_{H}$ is the binding energy for one hydrogen bond. For water
$E_{H}\sim10\, kJ/mol$ and $\alpha\sim100$ CGS, which is consistent
with the macroscopic value $\alpha\approx70$ CGS.

In the presence of impurities the induced polarization of molecules
propagates to a certain distance inside the bulk of the liquid and
can cause a considerable interaction of hydrophobic objects at large
distances. To analyze it we proceed to linear response study of polarization-polarization
correlations in our model. At large distances the interaction of external
objects does not depend on the precise structure of interacting bodies
and takes universal form. The asymptotic expression for the interaction
potential can be obtained by observing the free energy change in a
system of point ($\alt R_{D}$) chargeless impurities located at the
positions $\mathbf{r}_{a}$ characterized by a single vector property
$\mathbf{j}_{a}$. The simplest form of the interaction potential
is \begin{equation}
\Omega_{ext}=\int(\mathbf{j,s})dV,\label{eq:strongcoupling}\end{equation}
where $\mathbf{j}(\mathbf{r})=\sum_{a}\mathbf{j}_{a}\delta(\mathbf{r}-\mathbf{r}_{a})$.
For a specific shape of an interacting object the value and the direction
of the vector $\mathbf{j}$ can not be found within the linear theory
and requires microscopic derivation with the help of either complete
model, or molecular dynamics calculation. %
\begin{figure}
\includegraphics[%
  width=0.90\columnwidth,
  keepaspectratio]{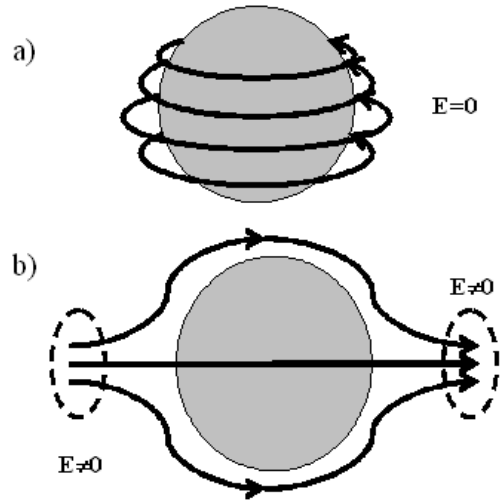}

\caption{Polarization $\mathbf{s}$ of the liquid around a nearly spherical
object. Forceless (a) and longitudinally polarized (b) liquid configurations.
\label{PolarizationNearSphere}}
\end{figure}

In the linear response approximation the polarization of the liquid
is \begin{equation}
s_{\alpha}(\mathbf{r})=-\sum_{\mathbf{k}\beta}G_{\alpha\beta}^{(s,s)}(\mathbf{k})j_{\beta}(\mathbf{k})\exp(i\mathbf{kr}),\label{Polarization}\end{equation}
where\begin{equation}
G_{\alpha\beta}^{(s,s)}=\frac{P_{0}^{-1}}{Ck^{2}+A}\left(\delta_{\alpha\beta}-\frac{4\pi k_{\alpha}k_{\beta}}{k^{2}}\frac{(k^{2}C^{\prime}/4\pi+1)}{(C+C^{\prime})k^{2}+A+4\pi}\right).\label{GreenFunction}\end{equation}
The interaction of the two impurities separated by a large distance
$\mathbf{R}=\mathbf{r}_{1}-\mathbf{r}_{2}$ is given by\[
\Omega_{L}(R)=-\sum_{k,\alpha,\beta}G_{\alpha\beta}^{(s,s)}(k)j_{\alpha}j_{\beta}=\]
\[
=\frac{1}{P_{0}^{2}}\frac{(\mathbf{j}_{1},\mathbf{j}_{2})-3(\mathbf{j}_{1},\hat{\mathbf{R}})(\mathbf{j}_{2},\hat{\mathbf{R}})}{A^{2}\epsilon R^{3}}.\]
This means that the two neutral impurities (\ref{eq:strongcoupling})
interact in unscreened fashion as two dipoles in a media with dielectric
constant $\epsilon$. The dipole moment $\mathbf{d}$ arises from
spontaneous polarization of water around the object: $\mathbf{d}=\mathbf{j}/(AP_{0})$.
We call such liquid configurations as longitudinally polarized states,
see e.g. Fig.\ref{PolarizationNearSphere}b for an example of water
polarization force lines around a nearly spherical body. For such
polarization type the non-vanishing polarization charge exists in
a liquid: $\rho_{P}=-P_{0}{\rm div}\mathbf{s}\neq0$. The induced
dipole moment around a microscopic object of the size $R_{0}\alt R_{D}$
can be estimated as follows: at small distances $r\alt R_{D}$ ($k\rightarrow\infty$),
the second term in the correlation function (\ref{GreenFunction})
can be neglected and $\mathbf{s}\sim\mathbf{j}/(P_{0}^{2}Cr)$. The
linear theory breaks on the surface of the object, $r=R_{0}$, where
$s\sim1$. Therefore $j\sim P_{0}^{2}CR_{0}$, $d\sim P_{0}CR_{0}/A$
and the energy of a pair of chargeless objects can be estimated as\begin{equation}
\Omega_{L}(R)\sim\frac{C^{2}R_{0}^{2}P_{0}^{2}}{AR^{3}}.\label{eq:Omega_SP}\end{equation}
This interaction is long-range and is completely due to the dipole-dipole
interaction of the water molecules. 

Longitudinal polarization of the liquid also contributes to the solvation
energy of a single impurity: \begin{equation}
\Omega_{solv}=-\sum_{k,\alpha,\beta}j_{\alpha}j_{\beta}G_{\alpha\beta}^{(s,s)}(k)=\Omega_{0}+\Omega_{A},\label{ActivationEnergy}\end{equation}
where $\Omega_{0}$ comes from the first term of Eq.(\ref{GreenFunction}),
formally diverges and can be estimated as $\Omega_{0}\sim j^{2}P_{0}^{2}k_{max}/C$,
where $k_{max}\sim R_{0}^{-1}$ is related to the size of the impurity.
The contribution from the second term is finite and is in fact the
energy (\ref{eq:Omega_polarization}) of the polarization electric
field: $\Omega_{A}\sim P_{0}^{2}C^{3/2}=50\, kJ/mol$. Since $\Omega_{0}\sim CR_{0}P_{0}^{2}$,
$\Omega_{A}\agt\Omega_{0}$ for impurities of sufficiently small size
$R_{0}\alt R_{D}$, the quantity $\Omega_{A}$ is large and can be
called as the activation energy of the longitudinally polarized configuration. 

\begin{figure}
\includegraphics[%
  width=0.90\columnwidth,
  keepaspectratio]{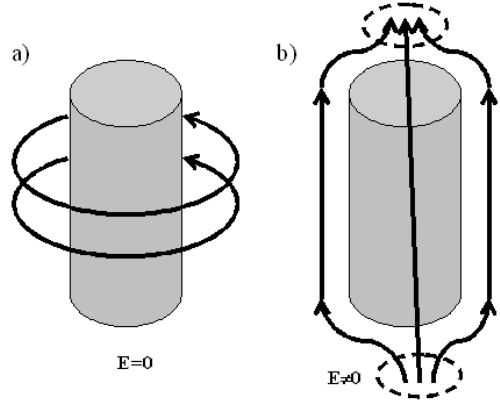}

\caption{Polarization of a liquid near a cylindrical hydrophobic surface.
Forceless (a) and longitudinally polarized (b) liquid configurations.
\label{PolarizationNearCylinder}}
\end{figure}

Large value of the activation energy is due to appearance of strong
electric fields next to polarized bodies. In fact there is a wide
class of liquid configurations with appreciably lower energies due
to the absence of polarization charge. This can happen in a system
of neutral bodies if a special, the {}``forceless'' (FL), state
of molecular dipoles is chosen:

\begin{eqnarray}
\nabla\mathbf{s} & = & 0,\mathbf{\; E}=0.\label{ForcelessREgion}\end{eqnarray}
Two examples of forceless water configurations around a sphere and
a cylinder are shown on Figs. \ref{PolarizationNearSphere}a, \ref{PolarizationNearCylinder}a.
The polarization vector in such states is similar to magnetic field
in magnetostatics, therefore the asymptotic form of the interaction
potential with with external point size water polarizing impurities
can be specified using the following form of the interaction potential:\begin{equation}
\Omega_{ext}=\int(\mathbf{J},\nabla\times\mathbf{s})dV.\label{eq:forcelesscoupling}\end{equation}
where $\mathbf{J}=\sum_{a}\delta(\mathbf{r}-\mathbf{r}_{a})\mathbf{J_{a}}$,
$\mathbf{r}_{a}$ is the position of an impurity, and $\mathbf{J}_{a}$
is a vector property of an impurity . The vector $\mathbf{J}_{a}$depends
on the details of the surface-water interactions and requires full
microscopic calculation for its determination. The energy of interacting
bodies is then:\[
\Omega_{FL}=\sum_{k,\alpha,\beta...}G_{\alpha\beta}^{(s,s)}(k)\epsilon_{\alpha\mu\nu}\epsilon_{\beta\mu^{\prime}\nu^{\prime}}k_{\mu}J_{\nu}k_{\mu^{\prime}}J_{\nu^{\prime}},\]
where $\epsilon_{\alpha\beta\gamma}$ is the totally antisymmetric
tensor. Using Eq.(\ref{GreenFunction}) we find that the contribution
of the second term vanishes. At intermediate distances $R\alt L_{T}$
\begin{equation}
\Omega_{FL}(\mathbf{R})=\frac{3(\mathbf{J}_{1},\hat{\mathbf{R}})(\mathbf{J}_{2},\hat{\mathbf{R}})-(\mathbf{J}_{1},\mathbf{J}_{2})}{4\pi P_{0}^{2}CR^{3}}\label{smallDistancies}\end{equation}
At larger distances $R\agt L_{T}$ the interaction vanishes exponentially:\begin{equation}
\Omega_{FL}(\mathbf{R})=\frac{(\mathbf{J}_{1},\hat{\mathbf{R}})(\mathbf{J}_{2},\hat{\mathbf{R}})-(\mathbf{J}_{1},\mathbf{J}_{2})}{4\pi P_{0}^{2}CRL_{T}^{2}}e^{-R/L_{T}}.\label{eq:Omega_FL}\end{equation}
The estimation for the vector $\mathbf{J}$ proceeds as follows: at
small distances $s\sim J/(P_{0}^{2}CR^{2})$ and has to be $s\sim1$
at the surface $R\sim R_{0}$ of the body. Hence, $\mathbf{J}_{1,2}=4\pi P_{0}^{2}CR_{1,2}^{2}\mathbf{n}_{1,2}$,
where $\mathbf{n}_{1,2}$ are the unit vectors connected with the
vorticity of polarization (see Figs. \ref{PolarizationNearSphere}a
and \ref{PolarizationNearCylinder}a) by the {}``right-screw rule'',
$R_{1,2}$ are the characteristic radii of the objects. The interaction
of a pair of chargeless impurities in a forceless water configuration
at small (\ref{smallDistancies}) and large (\ref{eq:Omega_FL}) distances
reads, correspondingly: \begin{equation}
\Omega_{FL}\sim\frac{CR_{0}^{4}P_{0}^{2}}{R^{3}}\label{SmallDistancies}\end{equation}
\begin{equation}
\Omega_{FL}\sim\frac{CR_{0}^{4}P_{0}^{2}}{RL_{T}^{2}}e^{-R/L_{T}}.\label{largeDistancies}\end{equation}
The exponential decay of the interaction at large distances is not
surprising since forceless configurations are also chargeless and
the net dipole momentum of the system is identically zero. Note, that
for a forceless configuration the polarization of the liquid is present
at distances $\lambda\sim L_{T}$ from the solute surface.

\begin{figure}
\includegraphics[%
  width=0.90\columnwidth,
  keepaspectratio]{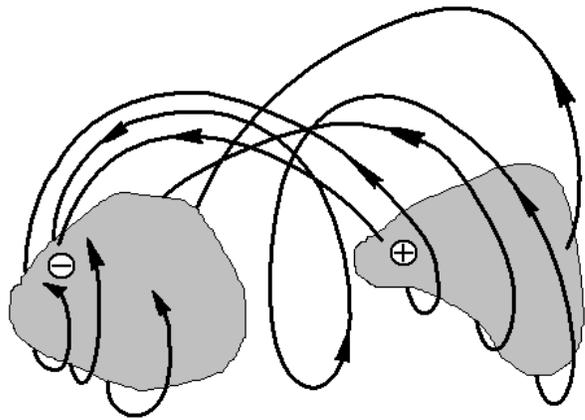}

\caption{Polarization of a liquid near two hydrophobic charged solutes.\label{TwoSolutes}}
\end{figure}

Eqs. (\ref{SmallDistancies}), (\ref{largeDistancies}) universally
characterize forceless interactions of small neutral objects. In linear
approximation (\ref{Polarization}) the polarization vector around
a pair of solutes located at points $\mathbf{\rho}_{1}$ and $\mathbf{\rho}_{2}$
is given by\begin{equation}
\mathbf{s}(\mathbf{r})\equiv\mathbf{s}_{FL}=R_{1}^{2}(\mathbf{n}_{1}\times\hat{\mathbf{r}}_{1})f(r_{1})+R_{2}^{2}(\mathbf{n}_{2}\times\hat{\mathbf{r}}_{2})f(r_{2}),\label{TwoBodies}\end{equation}
 where $f(r)=r^{-2}(1+r/L_{T})\exp(-r/L_{T})$. Here $\mathbf{r}_{1,2}=\mathbf{r}-\mathbf{\rho}_{1,2}$,
$\hat{\mathbf{r}}_{1,2}=\mathbf{r}_{1,2}/r_{1,2}$, and $R_{1,2}$
are the characteristic radial dimensions of the solute particles.
Eq. (\ref{TwoBodies}) is valid far from the objects ($r_{1,2}\gg R_{1,2}$).
If the solute bodies carry charges, than additional induced polarization
contribution appears: \begin{equation}
\mathbf{s=}\mathbf{s}_{FL}+\mathbf{s}_{P},\label{TotalPolarization}\end{equation}
\begin{equation}
\mathbf{s}_{P}\sim-\delta_{1}g(r_{1}/R_{D})\hat{\mathbf{r}}_{1}-\delta_{2}g(r_{2}/R_{D})\hat{\mathbf{r}}_{2},\label{PolarizationContributionFromBareCharges}\end{equation}
where $g(x)=x^{-2}\left[1-\left(1+x\right)\exp\left(-x\right)\right]$,
$\delta_{1,2}=q_{1,2}/q_{C}$. The water polarization is qualitatively
represented on the Fig. \ref{TwoSolutes}. Water polarization of this
type was obtained in MD simulations \cite{Higo} and called as {}``the
dipole-bridge between biomolecules''. The vortex-like structures
of polarization were observed that corresponds to the term $\mathbf{s}_{FL}$
in (\ref{TotalPolarization}). %

At intermediate distances between the interacting bodies both forceless
(FL) and longitudinally polarized (L) liquid configurations have very
similar interaction properties. Both $\Omega_{L}$ and $\Omega_{FL}$
are dipole-dipole interactions of polarized molecular dipoles. For
sufficiently small impurities we have $\Omega_{FL}/\Omega_{L}\sim R_{0}^{2}/R_{D}^{2}\ll1$.
This means that the interactions of small longitudinally polarized
impurities is stronger. On the other hand, the second term of Eq.
(\ref{GreenFunction}) does not also contribute to the solvation energy
of a single forceless impurity, which means that the forceless configurations
does not have the large activation energy contribution $\Omega_{A}\gg T$
in its free energy and hence, at least at large distances, thermodynamically
is more probable. At intermediate distances, $R\alt L_{T}(R_{0}^{2}/R_{D}L_{T})^{1/3}\ll L_{T}$,
$\Omega_{L}\agt\Omega_{A}$ and spontaneously polarized liquid configuration
with anti-parallel $\mathbf{j}$ possesses lower energy and may become
favorable. This means that anisotropic neutral molecules in a polar
liquid can acquire fairly strong orientation dependent interaction
caused by induced dipole moment of the surrounding water molecules.

\subsection{Interaction of macroscopic objects: parallel hydrophobic cylinders.}

The interaction potentials (\ref{eq:Omega_SP}) and (\ref{eq:Omega_FL})
give only the energies of point-like objects of size $R_{0}\ll R_{D}$.
In realistic calculations $R_{0}\agt R_{D}$(small organic molecules)
and may even be larger (proteins, lipids etc.). For larger objects,
a more detailed calculation has to be done. Consider first two parallel
cylinders of the same radii $R_{1,2}=R_{0}$, heights $H_{1,2}=H\gg R_{0}$,
placed at a distance $L\ll H$ from one another. The separation between
the objects is large, $L\gg R_{0}$, but the size $R_{0}$ is now
arbitrarily related to the water domain size $R_{D}$. At distances
large enough from the cylinder cores the superposition principle holds:
\begin{equation}
\mathbf{s}(\mathbf{r})\approx\mathbf{s}_{1}(\mathbf{r})+\mathbf{s}_{2}(\mathbf{r}),\label{Superposition Principle}\end{equation}
 where $\mathbf{s}_{i}$, are cylindrically symmetric forceless solutions
of model around isolated cylinders (see Fig.\ref{PolarizationNearCylinder}a):
\begin{equation}
\mathbf{s}_{i}(\mathbf{r})=\lambda_{i}\frac{R_{i}}{r_{i}}\beta_{i}\hat{\phi_{i}}.\label{SForCylinder}\end{equation}
Here $\lambda_{1,2}=\pm1$ stands for two possible directions of water
dipoles around a cylinder (the {}``topological charges''), $r_{i}$
are the distances from the point $\mathbf{r}$ to the axis of the
correspondent cylinder. The coefficients: $\beta_{1,2}\approx1/\left[1+\gamma R_{1,2}/R_{D}\right]$,
where $\gamma\sim1$ depends on the behaviour of the function $V(s^{2})$
for $s\sim1$. Formula (\ref{SForCylinder}) holds for $r_{i}\ll L_{T}$.
Since $\nabla\mathbf{s}\approx0$, the polarization electric field
vanishes. Substituting directly the solution (\ref{Superposition Principle})
into the energy density (\ref{TotalFreeEnergy}) we find the following
expression for the interaction potential: \begin{equation}
\Omega_{FL}(L)\approx2\pi AP_{0}^{2}\lambda_{1}\lambda_{2}\beta_{1}\beta_{2}R_{0}^{2}HK_{0}\left(\frac{L}{L_{T}}\right),\label{InteractionOfCylinders}\end{equation}
where $K_{0}$ is the Macdonald function. For small cylinders, $R\ll R_{D}$,
$\beta\approx1$ and the result (\ref{InteractionOfCylinders}) coincides
with that one can find by integrating the potential (\ref{eq:Omega_FL})
over the cylinders in the linear model. At sufficiently small distance
between the cylinders the interaction exceeds the activation energy
$\Omega_{A}$ and the longitudinally polarized configuration settles
(see Fig. \ref{PolarizationNearCylinder}b). Here $\nabla\mathbf{s}\neq0$
close to the ends of the cylinders, so that the energy of each of
the cylinders contains $\Omega_{A}\sim2\pi R_{0}^{2}\sqrt{C}\pi P_{0}^{2}$.
The superposition principle still holds and the energy of the cylinders
reads: \begin{equation}
\Omega_{L}(L)\approx2\pi CP_{0}^{2}\frac{\lambda_{1}\lambda_{2}}{\Lambda_{1}\Lambda_{2}}\beta_{1}\beta_{2}HK_{0}\left(\frac{L}{L_{T}}\right).\label{InteractionForLongitudinalPolarization}\end{equation}
Here $\Lambda_{1,2}=\log(L_{T}/R_{1,2})$. The result is again similar
to that one can find by integrating the interaction potential for
a spontaneously polarized configuration (\ref{eq:Omega_SP}) over
entire cylinders length.

Eqs.(\ref{Superposition Principle}) and (\ref{SForCylinder}) describe
an approximate forceless solution. Let us estimate the leading correction
$\Omega^{\prime}$ to the interaction energy due to the dipole-dipole
interactions of polarization charges (${\rm div\mathbf{s}}\neq0$).
Consider first two cylinders at a distance $L\ll L_{T}$. In this
case the polarization charges are concentrated around the cylinders
surface, $Q_{A,B}=-Q_{C,D}\sim P_{0}R_{0}$ in the sectors $A-D$
of Figure \ref{FigVortexAntivortex}. The dipole moment $D\sim P_{0}R_{0}L$
per unit length along the cylinders gives the following estimation
for the interaction energy\[
\Omega^{\prime}\sim\frac{1}{\epsilon(k\sim L^{-1})}\int dz_{1}dz_{2}\frac{\left[(\mathbf{D}_{1}\mathbf{D}_{2})-3(\mathbf{n}\mathbf{D}_{1})(\mathbf{n}\mathbf{D}_{2})\right]}{r_{12}^{3}}=\]
\begin{equation}
=\frac{2H}{\epsilon(k\sim L^{-1})L^{2}}\left[(\mathbf{D}_{1}\mathbf{D}_{2})-2D_{1x}D_{2x}\right]\label{DDInteractionOfCylinders}\end{equation}
 The integration in (\ref{DDInteractionOfCylinders}) is performed
along the axes $z_{1,2}$ of the cylinders. The vector $\mathbf{r}_{12}$
connects the points at the axes, $r_{12}=\sqrt{(z_{1}-z_{2})^{2}+L^{2}}$,
and $\mathbf{n=r}_{12}/r_{12}$. For a pair of cylinders of the same
size $\mathbf{D}_{1}=\mathbf{D}_{2}\equiv\mathbf{D}$ and\begin{equation}
\Omega^{\prime}\sim\frac{P_{0}^{2}HR_{0}^{2}(L^{2}+L_{T}^{2})}{\epsilon(L^{2}+R_{D}^{2})}.\label{DDResult}\end{equation}
 At larger distances, $L\agt L_{T}$, the solution (\ref{SForCylinder})
acquires an extra exponential factor $\exp(-r/L_{T})$ and \[
\Omega^{\prime}\sim\frac{P_{0}^{2}HR_{0}^{2}}{\epsilon}\exp(-2L/L_{T}).\]
Comparing this result with Eqs.(\ref{InteractionOfCylinders}) and
(\ref{DDResult}) we find:\[
\Omega^{\prime}(L)/\Omega_{FL}(L)\sim\exp(-L/L_{T})\le1.\]
This means that for sufficiently large distances the polarization
charges provide negligible corrections to the energy of a forceless
configuration.

\section{Solvent-solute interfaces. Phase transitions on boundaries.}

Consider now the structure of molecular polarization near a plane
liquid-to-vacuum or liquid-to-a body surface interface. Such a boundary
is hydrophobic and polarizes the liquid along the boundary plane $\Gamma$.
The polarization extends to the bulk of the liquid on a distance scale
$\lambda\sim L_{T}$ or $\lambda\sim R_{D}$ depending on electrostatic
properties of the boundary (see below). Since at larger distances
from the boundary the polarization of the liquid disappears, the polarization
itself is an effectively two-dimensional vector field. The Hamiltonian
for the surface polarization field $\mathbf{s}_{\Vert}$ can be obtained
by integration of (\ref{TotalFreeEnergy}) inside the bulk of the
liquid:\begin{equation}
\Omega_{S}\approx\frac{1}{2}M\int_{\Gamma}df\left(\nabla\theta\right)^{2}+\frac{1}{2}K\int_{\Gamma}df\left(\nabla\cdot\mathbf{S}\right)\phi(\mathbf{r})\label{Fluctuations}\end{equation}
 \begin{equation}
\phi(\mathbf{r})=\int_{\Gamma}df^{\prime}\frac{(\nabla^{\prime}\cdot\mathbf{S}^{\prime})}{\left|\mathbf{r}-\mathbf{r}^{\prime}\right|}.\label{Potential}\end{equation}
Here $\theta(x,y)$ is the angle characterizing the direction of the
unit vector $\mathbf{S}=\mathbf{s}_{\Vert}/s_{\Vert}=(\cos\theta,\sin\theta)$,
$\mathbf{r}=(x,y)$ are the coordinates along the surface, $\nabla=(\partial/\partial x,\partial/\partial y)$,
$\phi(\mathbf{r})$ is the electric potential at the point $\mathbf{r}$.
The constants $M\sim CP_{0}^{2}\lambda s_{\Vert}^{2}$, $K\sim\lambda M$
. Both terms in Eq. (\ref{Fluctuations}) originate from $\Omega_{H}$
and $\Omega_{dd}$ of Eq.(\ref{TotalFreeEnergy}), respectively. Note
that \begin{equation}
\nabla\cdot\mathbf{S}=-\sin\theta\cdot\theta_{x}+\cos\theta\cdot\theta_{y},\label{DivS}\end{equation}
where $\theta_{x}\equiv\partial\theta/\partial x$, $\theta_{y}\equiv\partial\theta/\partial y$.
Minimization of the functional (\ref{Fluctuations}) with respect
to the variations of $\theta(x,y)$ gives the following ''equilibrium
condition'':\begin{equation}
M\Delta\theta(x,y)+K(-\sin\theta\cdot\phi_{x}+\cos\theta\cdot\phi_{y})=0\label{EquationForTeta}\end{equation}
The simplest solutions of the selfconsistent equations (\ref{EquationForTeta}),
(\ref{Potential}), (\ref{DivS}) is the uniform polarization: $\theta(x,y)={\rm const}$
and \begin{equation}
\mathbf{S}={\rm const}\label{UniformPolarization}\end{equation}
\begin{figure}
\includegraphics[%
  width=0.90\columnwidth,
  keepaspectratio]{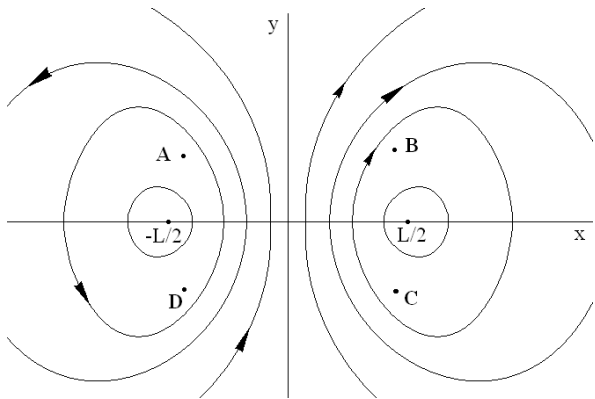}

\caption{Polarization configuration $\mathbf{S}(x,y)$ of surface water {}``vortex-antivortex
pair''.\label{FigVortexAntivortex} }
\end{figure}
 A more sophisticated solution describes a vortex state: $\theta(x,y)=m\arctan(y/x)$
of topological charge $q=m=0,\pm1,\pm2,...$. Both mentioned solutions
are forceless configurations and therefore $\lambda\sim L_{T}$. The
vortex core size is also $\sim L_{T}$. Fig.\ref{FigVortexAntivortex}
shows a more complicated polarization configuration, which is is a
vortex-antivortex pair. In this case $\nabla\cdot\mathbf{s}\ne0$
and thus $\lambda\sim R_{D}$ for all essential configurations with
vortex-antivortex pairs.

Let us neglect first the dipole-dipole interaction term $\Omega_{dd}$
in (\ref{Fluctuations}). In this case Eq.(\ref{Fluctuations}) coincides
with the Hamiltonian of XY 2D model. As it is well known from the
field theory \cite{Berezinskii,KosterlitzThouless}, thermal fluctuations
of the polarization field (\ref{Fluctuations}) can be mapped on to
the thermal dynamics of a gas of interacting vortices residing on
the surface. The thermal state of the liquid mainly consists of vortex-antivortex
pairs of minimum charge $|q|=1$. The energy of vortex-antivortex
configuration at large distance \begin{equation}
L\gg\lambda\label{LargeDistancies}\end{equation}
 between their cores is given by\begin{equation}
\Omega(L)\approx\Omega_{1}+\pi M\log\left(L/\lambda\right),\label{BKTEnergy}\end{equation}
where $\Omega_{1}\sim M$. We observe that the {}``hydrogen-bonding''
term $\Omega_{H}$ leads to appearance of the logarithmic attraction
of vortex and antivortex. At temperature %
\begin{figure}
\includegraphics[%
  width=0.90\columnwidth,
  keepaspectratio]{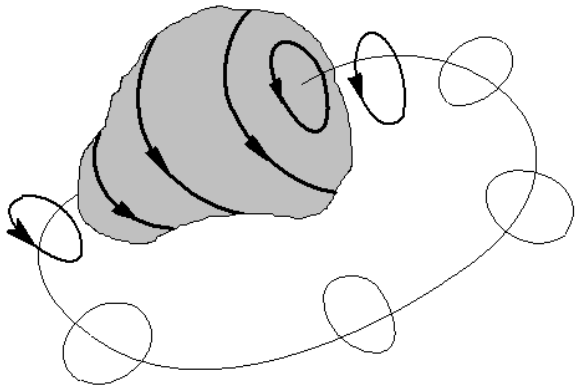}

\caption{Excitation of water with one vortex-antivortex pair pinned on a surface
of hydrophobic solute.\label{VortexAntivortexPair} }
\end{figure}
 \begin{equation}
T_{BKT}=\frac{1}{2}\pi M\sim\pi CP_{0}^{2}\lambda,\label{TBKT}\end{equation}
the topological Berezinskii-Kosterlitz-Thouless transition \cite{Berezinskii,KosterlitzThouless}
occurs: bound vortex-antivortex pairs dissociate and form the vortex
{}``plasma''. According to Onsager arguments \cite{Onzager,Kirkwood,Frenkel}
$\epsilon\sim P_{0}^{2}R_{D}^{3}/T$. On a hydrophobic surface $s_{\Vert}^{2}\sim1$,
and \begin{equation}
T_{BKT}\sim\epsilon T\gg T.\label{BKTforHydrophobic}\end{equation}
This means that at any reasonable temperature the polarization field
remains ordered (the molecular dipoles are correlated at large distances,
see below).

The long-range dipole-dipole interaction described by the second term
in Eq.(\ref{Fluctuations}) essentially alters the character of BKT
transition \cite{MaierSchwabl}. As is clear from Fig.\ref{FigVortexAntivortex}
the polarization vector $\mathbf{s}_{1}$ produced by the left vortex
encounters other vortex at the region next to the point $C$ and thus
induces the polarization density $\rho_{P}(C)>0$. At large distances
(\ref{LargeDistancies}) only one scale $L$ is essential: $\left|\mathbf{r}_{C}-\mathbf{r}_{D}\right|\sim\left|\mathbf{r}_{C}-\mathbf{r}_{B}\right|\sim L$,
therefore $\rho_{P}(C)\sim s_{\Vert}P_{0}/L$. The polarization charges
in each of the sectors equal: $Q_{C}=Q_{D}=-Q_{A}=-Q_{B}\sim\rho_{P}(C)\lambda L^{2}\sim s_{\Vert}P_{0}\lambda L$.
Roughly speaking the polarization charge pattern can be approximated
by the charge distributions of two equal downward oriented dipole
moments $D\sim Q_{C}L\sim s_{\Vert}P_{0}\lambda L^{2}$ placed at
a distance $\sim L$ one from another. Simple estimations give $\Omega_{dd}\sim D^{2}/L^{3}\sim KL$
and \begin{equation}
\Omega(L)\approx\Omega_{1}+\pi M\log\left(L/\lambda\right)+\ \beta KL,\label{MaierEnergy}\end{equation}
with $\beta\sim1$, instead of (\ref{BKTEnergy}). In other words
the dipole-dipole interaction of vortex with antivortex gives additional
attraction term with a linear dependence on $L$. The following remark
is necessary to explain this interaction. Indeed, the charge distribution
on the left side of Fig. \ref{FigVortexAntivortex} is a {}``mirror
reflection'' of that on the right side. Such charge distributions
repel in vacuum. In the case of a polar liquid interface the polarization
charges $Q_{C}$ increase as $L$ increases so that $Q_{C}+Q_{B}=0$
at any $L$. It means that in order to move apart a vortex and antivortex
pair, one should produce additional work to polarize the water molecules.
The Mayer-Schwabl phase transition occurs at \begin{equation}
T=T_{MS}=4T_{BKT}.\label{MayerTemperature}\end{equation}
 The phase transition is analogous to that for quark-gluon plasma
formation \cite{QuarkGluonPlasma}: unbound vortices are formed at
$T>T_{MS}$. To establish this fact one has to go beyond the approximation
of pair interaction of vortices (\ref{MaierEnergy}). According to
(\ref{MaierEnergy}) the typical dimension of vortex-antivortex pair
is: $L_{P}\sim T/K$. The surface concentration of pairs is: $n_{P}\sim\lambda^{-2}\exp(-\Omega_{1}/T)$.
At $n_{P}L_{P}^{2}\sim1$ we have a plasma-like quasineutral system,
in which topological charges of vortices and antivotices are mutually
compensated. Using our estimations for the parameters $K$, $\Omega_{1}$
and $\lambda$ we conclude that at $T>T_{MS}$ the interaction (\ref{MaierEnergy})
between two vortices is modified and the phase transition occurs.
In fact, Eq. (\ref{MaierEnergy}) between describes the interaction
of vortices only at $T\alt T_{MS}$ . At higher temperatures above
the phase transition point we can use a set of standard Debye arguments
as in linear approximation of screening in plasma (see e.g. \cite{LandLifshtiz}).
and we find the following mean-field effective interaction between
vortices of $Q_{T}$ placed at a distance $L$ from each other : \[
\Omega(L)=Q_{T}^{2}\int_{0}^{\infty}dpp\frac{pJ_{0}(pL)}{(p^{2}+\kappa^{2})p+\kappa^{2}p_{1}}.\]
Here $\kappa=\sqrt{2\pi n_{P}M/T}\sim1/\lambda$ and $p_{1}=K/(\pi M)\sim1/\lambda$.
Note that $\kappa\sim p_{1}$ at $T\sim T_{MS}$. At $T\gg T_{MS}$
$\kappa\gg p_{1}$, and at small distances, $L\ll1/p_{1}$, the the
interaction is logarithmic: $\Omega(L)\approx\pi MK_{0}(\kappa L)+\beta KL$
, which recovers Eq. (\ref{MaierEnergy}) at sufficiently small distances
$L\ll1/\kappa$. At large distances $L\gg1/p_{1}$ the screening switches
on and the interaction decays as $\Omega(L)\approx Q_{T}/(\kappa^{2}p_{1}L^{3})\sim1/L^{3}$. 

Eqs. (\ref{BKTforHydrophobic}) and (\ref{BKTforHydrophobic}) show
clearly that the phase transition can not be observed on a hydrophobic
surface. Nevertheless it may occur on hydrophilic surfaces, where
the energy of the liquid can be further lowered by extra-hydrogen
bonds involving the atoms on the surface of the solute. In a place
of such polar contact the vector $\mathbf{s}$ is almost perpendicular
to the liquid boundary: $s_{\Vert}^{2}\ll1$. The effective Hamiltonian
for the molecular dipole field on a hydrophilic surface still has
the form (\ref{Fluctuations}), though the phase transition temperature
is lower: $T_{MS}\sim CP_{0}^{2}\lambda s_{\Vert}^{2},$ $T_{MS}/T\sim\epsilon s_{\Vert}^{2}$.
The appearance of the additional small factor $s_{\Vert}^{2}$ can
decrease $T_{MS}$ drastically: the transition temperature may be
shifted to the room temperatures range already on a hydrophilic surface
with $s_{\Vert}\sim0.1$. 

At $T\agt T_{MS}$ the vortices dissociate and form vortex-antivortex
plasma, so that the order parameter $\mathbf{s}$ fluctuates strongly.
One may say that at small temperatures (or on a highly hydrophobic
boundary) the molecular dipoles fluctuations are small and the molecular
directions field is highly ordered on large scales. In other words
global network of hydrogen bonds exist on the surface of the solute
body. As the polar interactions of the solute object and the surrounding
liquid increase, the the phase transition temperature gets lower and
the global polarization order, and hence the hydrogen bonds network,
disintegrates. Moreover the hydrophilic regions on the surface of
the solute may pin the vortices, giving rise to a peculiar sophisticated
polarization shapes (see Fig. (\ref{VortexAntivortexPair}) and the
discussion in \cite{Higo}). 

We argue that the above discussed phase transition is directly related
to the disruption of the global hydrogen bonds network in the course
of 2D percolation phase transition observed in the Molecular Dynamics
simulation of hydration water absorbed on the surface of a solute
\cite{Oleinikova1,Oleinikova2,Oleinikova3}. In our model this result
can be explained with the help of classical field theory considerations.
At $T>T_{MS}$ a solute boundary hosts a gas of independent vortices
of different {}``circulation'' resulting in fast decay of correlation
between dipole moments of distant water molecules: $D(\mathbf{r})=\left\langle \mathbf{d}(0)\mathbf{d}(\mathbf{r})\right\rangle \sim\exp\left(-r/r_{C}\right)$
\cite{MaierSchwabl}. At $T<T_{MS},$ the vortices annihilate the
and long-range correlations set in: the ferromagnetic-like ordering
of hydration water molecules dipole moments takes place. It means
that in the considered case of polar liquids the boundary layer of
liquid is a ferroelectric film. Such correlated thermal states can
be seen as global network of hydrogen bonds at the surface. Note that
in our model the value of transition temperature $T_{MS}$ is proportional
to the width of polarized layer $\lambda$, which is in accordance
with the dependence of percolation transition temperatures obtained
in \cite{Oleinikova1,Oleinikova2,Oleinikova3} as a function the hydration
shell width $D$ (of course, it is reasonable to suppose that $D\sim\lambda$).%
\begin{figure}
\includegraphics[%
  width=0.90\columnwidth,
  keepaspectratio]{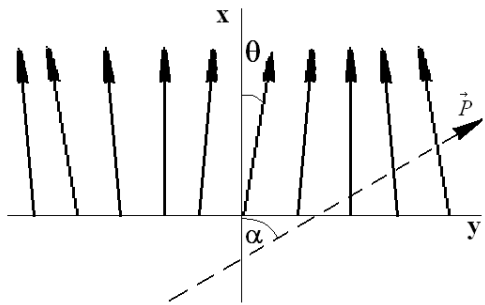}

\caption{Ferroelectric wave propagates in a film of hydration water.\label{ferroelectricwave}}
\end{figure}

The appearance of ferroelectric order can be used as a justification
for the mean field picture employed throughout the Letter. This fact
does not contradict with the Peierls-Mermin theorem \cite{Peierls,Mermin}
on the absence of ordering in a 2D systems due to new effect: the
long-range dipole-dipole term in (\ref{Fluctuations}). To illustrate
this point consider collective vibrations of molecular dipole moments
in a ferroelectric hydration water layer (Fig.\ref{ferroelectricwave})
with molecular dipoles arranged along $x$ axis. The kinetic energy
of a water molecule is $I\dot{\theta}^{2}/2$, where $I$ is the molecular
moment of inertia. The dispersion law $\omega=\omega(\mathbf{p})$
of the ferroelectric wave of a wave vector $\mathbf{p}$ can be established
by the minimization of the action $S=\int Ldt$, where the Lagrangian
$L$ reads\[
L=\int df\frac{1}{2}N\dot{\theta}^{2}-\Omega_{S}.\]
Here $N=\lambda nI$ and the interaction potential $\Omega_{S}$ is
defined by (\ref{Fluctuations}). The equation of motion\[
-N\ddot{\theta}+M\Delta\theta(x,y)+K(-\sin\theta\cdot\phi_{x}+\cos\theta\cdot\phi_{y})=0\]
can be expanded in powers of small $\left|\theta\right|\ll1$ so that
the dispersion relation is:\begin{equation}
\omega(\mathbf{p})=\sqrt{\frac{Mp^{2}}{N}+\frac{2\pi K}{N}p\sin^{2}\alpha}.\label{DispersionLaw}\end{equation}
Here $\alpha$ is the angle between vector $\mathbf{p}$ and $x$
axis. Consider the Peierls-Mermin argument applied for a 2D atomic
crystalline lattice. Mean squared displacements of the atoms from
the equilibrium positions in the lattice are proportional to the integral
$J=\int d^{2}p/\omega^{2}(\mathbf{p})$. Short-range interactions
of atoms give the phonon excitations with $\omega\sim p$. In this
case $J$ diverges at small values of $p$, which corresponds to fluctuations
with large wave lengths. The integral with the dispersion relation
(\ref{DispersionLaw}) does not have the singularity at small $p$:\begin{equation}
J\sim\int_{0}^{2\pi}d\alpha\int_{0}^{\infty}dp/(Mp+2\pi K\sin^{2}\alpha)\sim\int_{0}^{\infty}dp/\sqrt{p},\label{IntegralJ}\end{equation}
which means that the long-range dipole-dipole together with the short-range
hydrogen bonding interaction both stabilize the ferroelectric ordering
of the film in agreement with \cite{MaierSchwabl}. Interestingly
both $\Omega_{H}$ and $\Omega_{dd}$ are equally important in this
phenomenon: long range order does not exist both in the film with
the short and the long range interactions separately: the integral
diverges either if $M=0$ (no short range hydrogen bonds in the model)
and if $K=0$ (no dipole-dipole interaction). This somewhat surprising
observation means that the short-range behavior of intermolecular
interactions determine the character of order at large scales. Physically
it is a consequence of the Earnshaw's instability of the system of
classical dipoles, which is stabilized by the H-bonds. Similar findings
are reported in \cite{Belobrov,Brankov,Zimmerman,Bedanov} where the
2D system of classical dipoles were considered on a few different
types of lattices. It was shown that the ground state of the lattice
system can switch from a ferromagnetic to antiferromagnetic state
depending on the lattice shape.

\section{Concluding remarks. }

Finally we compare the results of our vector model to the density
functional theory, which is another well known approach to hydrophobic
interactions calculations. It was pioneered in earlier works of van
Der Waals \cite{vdW}, works well for simple, non polar liquids with
$\epsilon\sim1$, does not focus on electric properties and uses the
free energy functional expressed in terms of the mean $n(\mathbf{r})=\langle\rho(\mathbf{r})\rangle$,
the total $\rho(\mathbf{r})$ density of the liquid, and the density
fluctuations $\omega(\mathbf{r})$ ($\langle\omega(\mathbf{r})\rangle=0$)
\cite{Chandler,Lum,Wolde,Rowlinson,Nechaev}. The free energy functional
for the mean density is a non-linear function of $n(\mathbf{r})$
and is characterized by the healing length $a_{mac}\sim7\textrm{Å}$
for water. The fluctuations $\omega(\mathbf{r})$ are described by
the pair correlation function taken from experiments or molecular
dynamics simulations and is characterized by the second, microscopic
scale, $a_{mic}\sim1\textrm{Å}$. If the interaction between the molecules
of a liquid is purely short range, the correlation function for the
fluctuations decays exponentially at large distances:\[
G=\langle\omega(0)\omega(\mathbf{r})\rangle\sim\frac{\exp(-r/a_{mic})}{r}.\]
The interaction potential $\Omega$ of a pair of point-like objects
is proportional to the correlation function, $\Omega(r)\sim G(r)\sim\exp(-r/a_{mic})$
and, due to the scalar nature of underlying theory, is not orientation
dependent. Technically speaking the model suggested in this paper
also describes a model system with two characteristic scales, say,
$a_{mac}\sim L_{T}$ and $a_{mic}\sim R_{D}$ and thus can share many
predictions with the density functional models. On the other hand,
as was shown above the long-range dipole-dipole interaction between
the molecules of the liquid leads to a long-range power-law tail in
the correlation function (\ref{GreenFunction}):\[
\langle s_{\alpha}s_{\beta}\rangle\sim\frac{1}{r^{3}},\, r\rightarrow\infty.\]
This property is inherently related with the large dielectric constant
and is directly responsible for the strong long-range orientation-dependent
interactions described in this Article. Strictly speaking the liquid
in our model is incompressible, the assumption obviously fails at
$r\sim a_{mic}$. The complete model should include both fluctuations
of the liquid density (scalar) and the polarization (vector). One
example of such a model of a polar liquid was studied in \cite{Ramirez}.
Molecules were considered as freely oriented, like in the Onzager's
theory \cite{Onzager,Kirkwood,Frenkel}. The approach proved to successfully
reproduce the results of molecular dynamics simulations for solvated
ions. Another route to include the effects of liquid polarization
and electrostatic forces in the scalar model is outlined in \cite{Nechaev}.
The advantage of vector models is clearly seen when the effects of
the molecular polarization need to be taken into account, e.g. in
a case of charged solutes or for liquid boundaries and interfaces.
No scalar model can be directly applied to hydrogen bonds network
formation studies and topological excitations and interactions both
in the bulk and on the solute boundaries. 

The proposed vector model is both physically rich, relatively simple,
and is not much more computationally demanding than continuous solvation
models used in computational biophysics (see, e.g. \cite{Karplus,Onufriev}).
In its linearized form it reproduces a few very important features
of molecular dynamics simulations, such as spontaneous liquid polarization
next to macroscopic surfaces, the vortex-like structures in the molecular
dipoles arrangement near soluted bodies. Strong interaction between
the liquid molecules selects forceless liquids configurations with
vanishing polarization electric field. This leads to a strong, long-range,
orientation dependent interaction of solvated bodies. The model may
serve well as a continuous water model to describe different operational
states of nanodevices in aqueous environments, as well as membrane
charge transfer and ion channels, where both the hydrophobic and charged
boundaries are equally important. 

Fast solvation models are crucial in drug discovery and biomolecules
modelling. The quality of solvation models is a limiting factor determining
the accuracy of the calculations in molecular docking and virtual
screening applications. The proposed model is employed in QUANTUM
drug discovery software for very accurate binding affinity predictions
for protein-ligand complexes \cite{qpharm}. QUANTUM uses the continuous
vector water model (\ref{TotalFreeEnergy}) for solvation energy predictions,
determination of protonation states of biomacromolecules, pKa calculations,
protein structure refinement (including mutagenesis) etc.


\begin{thebibliography}{10}
\bibitem{PieceOfIce}If a solute charge exceeds $q_{C}$ the linear model breaks down close
to the charged molecules and the adjacent water molecules may become
{}``frozen''. The discussion of this phenomena will be published
elsewhere.
\end{thebibliography}
\end{document}